\documentclass[aps,prb,showpacs,twocolumn,superscriptaddress,nofootinbib]{revtex4-1}
\usepackage{bm,color,amsmath,amssymb,mathrsfs,latexsym,graphicx,psfrag,mathtools}
\usepackage{graphicx}
\usepackage{bm}
\usepackage{color}
\usepackage{amsmath}
\usepackage{epsfig}
\usepackage{epstopdf}
\usepackage{comment}
\usepackage[bottom]{footmisc}

\newcommand{\bea}{\begin{eqnarray}}
\newcommand{\eea}{\end{eqnarray}}

\begin{document}

\bibliographystyle{unsrt}

\title{A New Thermodynamics Featuring the {\it Arrow in Time}}
\author{James P. Hurley, Department of Physics and Astronomy,\\
University of California Davis, Davis CA 95616 }
\date{\today}

\begin{abstract}
It is a truth universally acknowledged that all isolated macroscopic systems must be in want of ever greater disorder, with due apologies to Jane Austen for plagiarizing the opening line in her novel {\it Pride and Prejudice}. This common, everyday process, often referred to as the Arrow of Time, has not been able to find a comfortable home in the body of modern statistical physics. There is a hint of such a disordering arrow in the second law of traditional thermostatics but, properly stated, that law relates only to the static equilibrium state of confined, isolated macroscopic matter. We will attempt to enhance today's thermo{\it statics} with a true thermo{\it dynamics} in which the 
disordering arrow in time is shown to be responsible for all the order
-- creation we find in the universe -- including the existential order needed to sustain life.
\end{abstract}
\date{\today}
\maketitle

\section{Introduction}

Life's big question:
there are two books with the same title of {\it What is Life?} The first 
was published in 1944 by physicist and Nobelist Erwin Schr\"odinger, and 
the other in 2021 by biologist Paul Nurse, also a Nobelist. Nurse 
acknowledged his pilfering of Schr\"odinger's title\cite{note1}  
as follows: 
``Even the title of this book {\it What Is Life?} has been shamelessly 
stolen from a physicist, Erwin Schr\"odinger, who published an influential 
book of the same name in 1944. His main focus was on one important aspect 
of life: how living things maintained such impressive order\cite{note2}  and uniformity 
for generation after generation in a universe that is, according to the 
Second Law of Thermodynamics,\cite{note3} constantly moving towards a state of disorder 
and chaos. Schrödinger quite rightly saw this as a big question ...''
\\ 
So, the Big Question in understanding life is not so much {\it What is life?} but,  {\it How is life?}  
“How do living things maintain such impressive order ... in a universe that is ... constantly moving 
towards a state of [ever greater] disorder and chaos,” that is to say, compulsively follow the 
disordering arrow in time? It is an existential question.

\section{A new thermodynamics}
To answer this question we need to construct a new branch of statistical 
physics, a new thermodynamics. Properly stated,\cite{note4}  the traditional second law of thermostatics 
says only that:  
{\it There exists an entropy function, generally symbolized by the character S. It is a unique function of the extensive properties of the system (in a simple system those properties are the energy, volume, and particle number for each particle type) and is a maximum in the equilibrium state.} 

\vskip 1mm
This statement says almost\cite{note5} nothing about the 
statistical processes by which 
the equilibrium state is achieved -- despite all the rhetoric to the contrary. 
And so there is a need to broaden the current thermo{\it statics} to a thermo{\it dynamics} 
that recognizes the existing asymmetry in time in a ``universe that is 
constantly moving towards a state of disorder and chaos,'' and the existence 
of a statistical arrow in time that is found in all living organisms, indeed 
in all isolated macroscopic matter in the universe!
	But first, why focus on ``statistical processes''? 

Here is the brilliant James Clerk Maxwell on necessity of acquiring a 
second ``kind of knowledge'' when dealing with macroscopic 
matter:\cite{note6} 
``I think the most important effect of molecular science [macroscopic matter] on our way of thinking will be that it forces on our attention the distinction between two kinds of knowledge, what we may call for convenience, the Dynamical and the Statistical ... Now if the molecular theory of the constitution of bodies is true, {\it all our knowledge of [macroscopic] matter is of the statistical kind.'' [Italics mine.]}\\
In other words, in physics we count: 1, 2, 3, statistics.

If we are to understand the behavior of macroscopic matter we should avoid 
``Dynamical'' knowledge, the difficult, opaque, time-symmetric laws of 
particle physics, quantum mechanics, and cosmology, and focus instead on the
intuitive, time-asymmetric, transparent ``Statistical'' knowledge of
a new thermodynamics. 
	
To see this new field of  thermodynamics in action, consider the following 
``Dynamical'' experiment: A gas of identical, weakly interacting particles 
is initially confined to the left half of a box by a vertical partition at 
the mid-plane. There is a small hole in the partition, but it is initially 
plugged.
 
Now, remove the plug, thereby {\it passively} relaxing a constraint on the system. 
The gas begins to uniformly diffuse through the hole in the partition from 
the high density left half of the box to the low density right half. {\it Never}, 
at any time, is there even a single reverse, ordering flux from the right 
half to the left half. This continues for some time, but eventually all 
variation in the macroscopic features (pressure, density, temperature, 
viscosity, specific heat, etc.) cease; macroscopically a {\it stable} equilibrium 
state has been achieved, despite vigorous activity remaining at the 
microscopic level. 

Notice that there are two distinctly different statistical phenomena at play 
in that experiment, one, the {\it uniform, inexorable time-asymmetric diffusion} 
of the gas from high density left side to low density right side, and two, 
the rock-solid stability of the eventual, static equilibrium state. These 
are quantitative, reproduceable results, making them the stuff of science, 
the statistical science of macroscopic matter. 

We recognize the eventual equilibrium state as equivalent to the 
traditional second law of thermo{\it statics}, as well as a new time-asymmetric 
thermodynamics in which there is an ever-increasing entropy/disorder as 
a manifestation of what traditionally goes by the name {\it The Arrow in 
Time}. Generally speaking that law says:\cite{note7}
``The entropy/disorder of all isolated, macroscopic systems remains 
constant, or increases uniformly with the passage of time''.\cite{note8}

It is that disordering arrow in time that accounts for the ``Dynamics'' 
in this new thermodynamics, and makes explicit the asymmetry in time in 
the behavior of confined, isolated macroscopic matter, including but not 
limited to, all life. We will embrace this arrow as the fourth law in 
our new science of thermodynamics -- Nernst's law of thermostatics 
being the third law.

\section{The Order Creation Principle}
Unfortunately, the arrow in time, if it is to be directly useful, does not appear to be the answer to life's Big Question raised earlier, for that arrow seems to point in exactly the wrong direction: Life needs to overcome the natural tendency to follow the disordering arrow in time, and become ever-more ordered. 

Never mind the ``wrong direction.'' We will perversely exploit that misdirected arrow to our advantage in what we will call the: \\
{\it Order Creation Principle}.
In isolated, {\it multicomponent}, macroscopic systems it is possible for one component, 
{\it with ready access to a state of greater entropy/disorder}, to spin-off a fraction 
of its order (but not the whole of it) to another, lesser component, often mediated 
through the use of facilitating tools, demanding only that the net entropy/disorder 
must increase. 
\vskip 1mm \noindent
{\it Nota bene}: \\
$\bullet$ The physics behind the principle is straightforward and well-known. But, given how crucial its application is in our new thermodynamics, we need to be able to address it by name, and so 
we have christened it the {\it Order Creation Principle.}\\
$\bullet$ For this principle to work, one component must be {\it ordered}, that is 
to say have a {\it ready access to a macrostate of greater entropy/disorder}, for it is 
that access that provides the statistically compelling {\it agency for change}. \\
$\bullet$ Note also that it is that {\it access} of macroscopic matter to greater 
entropy/disorder, and not, as is so often assumed, the energy that provides the agency 
for change.\cite{note9} 
Energy is static and without direction. A one pound stick of dynamite has the same energy 
as a one pound brick (remember, $E=mc^2$), but clearly the dynamite provides a greater, indeed explosive, agency for change; energy is not energetic. The most we can say about it is that it is the {\it thing that is conserved} -- good luck with just that. On the other hand, ordered macroscopic matter, with {\it ready access to macrostates of greater entropy/disorder}, relentlessly follows the disordering arrow in time and thereby provides the agency {\it }and direction for change.\\
$\bullet$ We also see that there is no ``energy crisis,'' only a crisis in the availability of {\it macroscopic systems with ready access to a state of greater entropy/disorder} (in the given environment\cite{note10}), 
that is to say, by definition, a lack of available {\it ordered} systems, fossil fuels for example. (Fossil fuels, being fossils, were ordered ages ago, being once themselves alive.) Unconstrained, these 
ordered macroscopic systems are the drivers (agency) of change as they autonomously {\it pursue} the disordering demanded by the arrow in time. Energy is germane only in the absence.\cite{note11}
\vskip 1mm 	
As an everyday example of the order creation principle, consider as a two component system, one, the ordered sunshine streaming from the sun, and two, the Earth's atmosphere. Sunshine, the ordered direction of steaming photons coming from the sun, with ready access to becoming disordered as they are differentially absorbed by the variously shaded portions of the Earth’s surface. Those dark patches are warmed and the light patches less so, and that variation in the local heating generates surface winds (ordered flow of air). And so, 
\begin{enumerate}
\item Sunshine as the component with ready access to greater disorder,
\item atmospheric winds as the recipient of the order borrowed from sunshine,
\item and the nonuniform shading of the Earth as the tool to make it happen. 
\end{enumerate}
One form of order, {\it sunshine}, has been converted to another, lesser forms, winds 
in the atmosphere, using the variously shaded Earth as the tool to {\it borrow} a 
fraction of the sunshine's order. Winds in the atmosphere borrow their order 
from sunlight: the  {\it Order Creation Principle} in full display.

Note also: This suggests the following universal principle: 
{\it The Order Creation Principle provides us with the mechanism by which virtually 
all order is created, ``even in a universe that is constantly moving towards a state of 
disorder and chaos!}''
As we will see, that is the answer to the Big Question raised by Schr\"odinger and Nurse.

\section{Can macroscopic matter self-order?}
But there are apostates among us. 
\noindent 
\begin{enumerate}
\item {\it First apostate}. We read in a popular 
science book,\\
``The universe is clearly an increasingly structured place ... The primordial ball of fire has cooled and condensed from a hot, amorphous mass into elements, stars, planets, and people. What we see in the universe is increasing in order, not disorder.''
\item {\it Second apostate}. Murray Gell-Mann, in his brilliantly imaginative book, 
{\it The Quark and the Jaguar}, pg. 230, writes   
``As the entropy of the universe increases, self-organization can produce local order, 
as in the [formation of stars], arms of a spiral galaxy or the multiplicity of symmetrical 
forms of snowflakes.''
\item {\it Third apostate}. There was a time when it was assumed that the universe did not have sufficient energy to continue to expand forever. There would come a time when the expansion would slow, stop, reverse itself and begin to contract back toward its birthplace in the Big Bang. It was called the Big Crunch. 
\end{enumerate}
It was further assumed that since the universe was self-disordered on expansion, it would self-order on contraction. It was furthermore assumed that the direction of time was irrevocably coupled with the direction of increasing entropy/disorder. Here is Stephen Hawking in his best-selling book, {\it A Short History of Time}, 
where he speaks of time reversal during the crunch: `
``At first, I believed that disorder would decrease when the universe recollapsed 
[in the Big Crunch]. This was because I thought that the universe had to return to a 
smooth and ordered state when it became small again. This would mean that the contracting 
phase would be like the time reverse of the expanding phase. People in the contracting 
phase would live their lives backward: they would die before they were born and get 
younger as the universe contracted.'' 
[Note: later he recanted.]

{\it Assessment of the apostates.} In these examples of so-called ``self-ordering'' it was assumed that the order in question was a function of the spatial configuration of macroscopic matter alone; if spatial expansion was disordering, it was assumed that spatial contraction during the crunch must therefore be ordering. 

\subsection{Other sources of disorder}
There are autonomous macroscopic disordering processes in which there is no {\it spatial} 
reconfiguration to speak of. Consider the example of two blocks, one hot and one cold and 
allow them to come together and exchange heat. In that exchange the hot block is cooled, 
and thereby ordered, and the cold block is heated, and thereby disordered. But what 
exactly is it that is ordered or disordered? There are no moving parts to speak of! 
There is no spatial restructuring whatever! So what is it that is ordered or disordered? 

Since the heat exchange happens autonomously in macroscopic matter, it must follow the disordering arrow. To verify this just do the thermodynamics. Let $dQ$ be a differential amount of heat exchanged at any time during the heat exchange. The net change in entropy is given by\cite{note12}
\bea 
dS &=& dQ/T_{cold} + (-dQ)/T_{hot} \nonumber \\
   &=& dQ (1/T_{cold} -1/T_{hot}) > 0,
\eea
where $dQ$ is the amount of heat add to a block, $dQ$ is negative in the hot block. The change in entropy is therefore greater than zero since $1/T_{cold}$ is greater than $1/T_{hot}$.

And so we have a proof of concept:   
{\it it is possible, in a closed, isolated, macroscopic system, to create order autonomously, even in the face of the disordering arrow in time; the arrow demanding only that the net entropy/disorder must increase.} 
But, again, there is virtually no spatial change in the particles of either block. There is no change in positional order. So what is it that is disordered? 

The missing component is most easily understood by the application of the 
{\it Kinetic Theory of Gases}\cite{note13} to an ideal gas in which the microstate is 
characterized by the position {\it and} the {\it velocity of each particle, and the 
positions and velocities are to be counted independently}. The total disorder 
becomes the {\it product} of the two, the positional and kinetic disorders. 

On contraction of a gas cloud to form a star, it would be true, as Gell-Mann 
said, there would be a {\it positional} ordering on contraction of an 
interstellar ideal gas cloud, but there would be an even greater {\it kinetic} 
disordering in the conversion of ordered gravitational energy to disordered 
thermal energy. Suns are hot; suns are kinetically disordered as they form, 
more so than they are positionally ordered. 

Even that well-structured snowflake of Gell-Mann is more kinetically disordered 
in the conversion of bonding energy in the crystal to thermal energy, than it 
was ordered in the spatial ordering in the formation of the structured crystal. 
Overall the total disorder increases. Hard as it may be to imagine, that 
exquisitely structured crystalline snowflake is the most disordered possible 
state of those water molecules under the given conditions of pressure and 
temperature. We know this to be true if only because that formation was 
autonomous, following the arrow in time to ever greater 
entropy/disorder/probability.

In general the nature of the macrostate is complex, and cannot be characterized as either spatial or kinetic, but some complicated mix of the two. 

\section{Laws of Thermodynamics}
\subsection{The authority by which the laws of thermodynamics speak to us}
It is instructive to compare the authority by which these two aspects of this new 
thermodynamics: one,  the existence of a stable equilibrium state and two, the 
arrow in time leading to that stable end state—speaks to us.
First we will focus on the authority by which the equilibrium macrostate speaks, 
the eventual lack of variance in all macroscopic features in the equilibrium state 
of isolated macroscopic matter. (We will discuss the authority of arrow presently.)

In our observations of macroscopic matter we do not see individual particles, we (or our instruments), following Maxwell’s advice, take statistical averages over a multitude of signals sent or reflected by a multitude of particles. In any average there is always a margin of error. Mathematical statistics has something important to say about such error; it is the smoothing effect of the law of large numbers: 
{\it The variance $\sigma$ in the average of N random and independent variables 
(observables) equals $\sigma_1/\sqrt{N}$, where $\sigma_1$ is the variance in 
just one member of group.} 
Schr\"odinger, in his book {\it What is life?} calls this ``the $\sqrt{N}$ law,'' 
and views it as a measure of the {\it stability} of the equilibrium state. It is 
that stability that allows us to meaningfully quantify macroscopic matter. 
We say the temperature T=98.0$^{\circ}$, and not T=$98.6^{\circ}\pm 16^{\circ}$.

To the extent that the signals we average in our observations of a large number
$N$ of particles in macroscopic matter are random and independent, the variance 
in the averages we take in those observations will be on the order of $1/\sqrt{N}$, 
or, where $N$ is on the order of Avogadro's number, $6.02\times 10^{23}$, 
the variance will be on the order of   
$10^{-12}$. And so we rewrite the second law of the old 
thermostatics:
{\it That variance of $10^{-12}$ is the authority with which the 
equilibrium state of macroscopic matter speaks to us, despite vigorous 
activity at the microscopic level.}

The high degree of that stability of the equilibrium state may seem strange at first, given the wild fluctuations at the microscopic level. The physical basis for this stability is, however, the smoothing effect of the law of large numbers, itself, not entirely intuitive! It is, nevertheless, an example of the near universal aphorism:
{\it There are behaviors in crowds that are not foreseen in the elements that make up the crowd}. ` 
In the human domain, the crowd storms the Bastille while the individual writes a letter to the editor. 
	
Next, as promised, we ask with what authority does the arrow in time speak to us 
in this new thermodynamics? What assurance do we have that the entropy/disorder doesn't 
sometimes increase and sometimes decrease, staggering about haphazardly as the system 
lurches toward the eventual trap of the highly stable equilibrium state?
Unfortunately, there is no law of large numbers for system 
trajectories, the way there was for the stable equilibrium state above, 
that $1/\sqrt{N}$ law. But it is possible to establish a reasonable 
estimate for the inexorable uniformity of the increase, again appealing 
to the Kinetic Theory of Gases. 

Earlier we considered a box in which the particles of gas were unevenly distributed between the left and right halves of a box. If the particles interact only weakly we may apply the 
Kinetic Theory of Gases\cite{note13} to quantitatively study the diffusion from the 
high density left half of the box to the low density right half. It allows us to 
treat the particles as a random collection of atomic or molecular particles 
stochastically moving about and thus calculate the fraction of ordering and 
disordering trajectories, that is to say, the fraction of the time in which the 
particle flux is from the high density left side to the low density right side 
(thus disordering) and, separately, the fraction of the time in which the flux is 
from the low density right side to the high density right side (thus 
ordering).\cite{note14} It is a tedious calculation; no need to 
reproduce it here, but in an ideal gas, initially at STP, we have 
found\cite{note15}  the chance of there being even a single ordering trajectory 
among the profusion of disordering trajectories at any point during that evolution 
toward that final equilibrium state to be on the order of 
one chance in $10^{10^{18}}$, and so the variance in the trajectory of 
macroscopic matter is on the order of $10^{-10^{18}}$, providing us with 
the following general principle:  
{\it That variance of $10^{-10^{18}}$ in the trajectory of macroscopic 
matter represents the authority with which the arrow in time speaks to us.}

We will settle on this figure. It isn't definitive, or all inclusive, 
but the odds of there being a variance of even 1 chance in $10^{10^{18}}$ 
is small, exceedingly small, and that’s all that really matters; 
corrections on the order of $10^{10}$ don't matter much. And so, we 
conclude that macroscopic matter doesn’t just wander about aimlessly 
until it accidently stumbles into that stable equilibrium state; it 
pursues that equilibrium state {\it relentlessly}! It is the relentless 
provider, the agency, for change in the nonequilibrium state of 
isolated macroscopic matter.\cite{note16}

Once again we might observe: there are behaviors, trajectory 
behaviors this time, in crowds that are not foreseen in the elements 
that make up the crowd. Stanford professor of physics Leonard Susskind, 
was himself puzzled by this arrow. In his Messenger Lecture titled, 
``Boltzmann’s Arrow of 
Time,''\cite{note17} he lamented: ``The arrow of time is a fact of nature and needs an explanation.''  That variance of 10$^{-10^{18}}$ in the 
disordering trajectories of macroscopic matter provides the needed explanation. 

To understand where such numbers as $10^{-10^{18}}$ come from, 
here is the expression\cite{note18} for the fraction of ordering 
trajectories, $\frac{dn_L}{dt}>0$), that is, trajectories in which 
there are $n_L$ particles in left half of the box, $n_R$ particles 
in the right half, $n_L$ greater than $n_R$, and yet $n_L$ is 
increasing during the scaled time of $\Delta\tau$:
\bea
F\bigl(n_L,\frac{dn_L}{dt}>0\bigr)=
   \frac{exp[-(\sqrt{n_L}-\sqrt{n_R})^2\Delta\tau]}
      {\sqrt{\pi} (n_Ln_R)^{1/4} \Delta\tau ln(n_L/n_R)}
\eea
Note that the numerator of that fraction on the right side is exponential in the square root of the particle numbers, which are themselves exponential, on the order of Avogadro’s number $6.02\times 10^{23}$. Thus numbers on the order 
of the $exp(10^{18})\sim 10^{{10}^{18}}$ are not surprising.

So, now, at long last, given the authority with which the arrow in time 
speaks, we must acknowledge that we do not play fast and loose with the 
macroscopic future. Novelists may dabble in the fantasy of a {\it Connecticut 
Yankee in King Arthur's Court} foretelling an eclipse. Or cosmologists may allow individual particles to make round trips on so-called {\it closed 
time-like curves}, or sneak single file through wormholes, allowing 
individual particles to converse briefly with the ``past'', but these 
fabulists are not allowed their whimsy in crowds! Crowds do not sneak 
and will not be marshaled in single file. They pursue that rigidly 
stable equilibrium state with a vengeance! 

\subsection{Implications of that authority} 
Given the authority of these two statistical laws, one, the stable equilibrium state and two, the relentless disordering arrow in time in a nonequilibrium state, with variances of $10^{-12}$ and $10^{{-10}^{18}}$ respectively, it is worthwhile to pause for a moment to appreciate some of the related implications.

\subsubsection{First, the stability of the equilibrium state} 
	The stability of the equilibrium state explains why the Mona Lisa's smile, enigmatic though it may be, is, at the very least, not a blur. That smile is all about the average our senses take over the very large number of signals sent, or better, reflected, by the very many molecules within the pigments of paint that make up the smile, and, being an average over a great many, 
roughly speaking, random and independent elements, the law of large 
numbers, that $\sqrt{N}$ law, tells us that there will be little or 
no variance in that average. And so we understand why the smile is not a 
blur. (My apologies for the crudeness of this application to such an 
iconic, artistic masterpiece as the Mona Lisa painting. Keats was right 
to lament in his poem {\it Lamia}, ``Do not all charms fly at the mere 
touch of cold philosophy?'') 

This same stability applies to the air in the room you now occupy, the lack of micro-air currents, and the lack of variance in the chemical elements in the chemist’s test tube. The margin of error in the averages we take with macroscopic 
matter at equilibrium is essentially zero, making macroscopic matter 
worthy of inclusion in the body of {\it quantitative} statistical 
science, {\it i.e.} traditional thermostatics.

\subsubsection{The stability of the macroscopic trajectory} 
The near zero variance in the arrow of time puts to rest the impression 
left by Richard Feynman in an engaging talk he gave to a group of 
Cornell students on {\it The Distinction of Past and Future.} 
He began his discussion of that distinction by showing how planetary 
bodies (they eventually became particles in his talk) might move 
about in the gravitational field of the Sun: 
``If you have a lot of particles [planets] doing this,'' Feynman said, 
winding one hand about the other, ``and then reverse the speeds,'' 
now unwinding the rotation of his hands, ``they will completely undo 
what they did before.''

It seems to imply that for any macroscopic system with a trajectory evolving one way in time, there will be a velocity-reversed trajectory that evolves in precisely the other way; for every disordering trajectory there will be a compensating ordering trajectory. Of course that is not true; there is a very great surplus of disordering trajectories -- with a variance of $10^{-10^{18}}$. The flaw in that argument is that Feynman's velocity-revered trajectory is not a part of the same system in which the original trajectory began. In a Feynman-like to-and-fro of flux between the two halves of a box, one may have started with a ``to'' in which there is an 80\%/20\% distribution of particles between the two halves of the partitioned box, but it returns as a ``fro'' in which the distribution is then 79\%/21\%. Not the same.

The issue is whether for every disordering trajectory, in the 80\%/20\% 
distribution, there is an ordering trajectory. There is not! There is a 
superabundance of disordering trajectories -- with a variance on the 
order of 10$^{-10^{18}}$ in macroscopic matter.  It’s as simple as that! 
This so-called time-asymmetry paradox is dealt with at length -— perhaps to excess -— in Hurley (2019).\cite{note7} 

\subsubsection{The arrow in time verses the arrow of time}
Before we pursue these applications further, we must interject an 
important disclaimer. There is more than one arrow in time. 
Roger Penrose\cite{note19} 
 was able to tease out seven variants, seven different arrows. We have 
not been willing to venture into that treacherous mine field of 
multiple arrows, and so we have always been careful to confine ourselves 
to the traditional thermodynamic arrow, the arrow that is conventionally, 
but unjustly, associated with the second law of thermostatics, traditional 
thermodynamics, the arrow in which the entropy/disorder of confined, 
macroscopic matter increases uniformly with time. 

	But that thermodynamic arrow defines what matter does, not what 
time does; time marches to its own drummer. The disordering arrow has 
little or nothing to do with time, but everything to do with the 
statistical nature of a great many particles randomly moving about, 
{\it en masse}, always evolving toward statistically-compelling more 
probable macrostates -- with a variance in that pursuit of 
$10^{-10^{18}}$. And so we have always spoken of  the ``arrow in time,'' 
and never the customary ``arrow of time,'' despite all the fanciful 
rhetoric to the contrary. 

	Some have even said that the increase in entropy with time is 
a ``property of time itself.'' Oh, how I wish I were able to find some 
meaning in that fabled statement: ``...macroscopic disordering being a 
property of time itself!'' But, outside cosmology, I can't even begin 
to imagine what that might mean; it is truly mind boggling. 

\subsubsection{Answering the Big Question} 
And now, to be specific, how exactly does the {\it Order Creation 
Principle} work in the living cell, how does the cell, in Schr\"odinger's 
language, ``continually draw neg-entropy [i.e. order] from the environment.''
We have shown how order can be created in one component of a 
multicomponent system when there is an even greater disordering in some 
other component. But how do we know, in the cell, that the other 
component is more disordered?\cite{note20}

	We have considered the case of the thermal heat exchange between 
two blocks, where we were able to verify the increase in the total 
disorder. But we now need an example that is more directly related to 
the Ling cell.
	
As an example of how that Big Question of Schr\"odinger and Nurse is 
answered in the cell, consider how order might be maintained in a 
well-run, working city. Consider as a two component system in that 
city, the local surface winds and the water in the city's water tower. 
The city's windmill (the tool) feeds on an ordered local surface winds 
to pump ground water up into the city's water tower, there to be stored 
to provide a source of order with ready access to a state of greater 
disorder for future city use.\cite{note21}  There are two 
ordered components, 
wind and water, and a windmill as the tool, to create that stored 
order in the city's water tower.\cite{note22}

That city provides us with a metaphor for how life answers the Big Question in the living cell. In the multicomponent living cell, think of food, apple juice perhaps, as the ingested ordered component with ready access to greater disorder, and the naturally evolved cellular organelles\cite{note23}  (endoplasmic reticulum, Golgi bodies, lysosomes, ribosomes, mitochondria, plastids, the cell membrane itself, and chloroplast in plant cells) as the tools that feed on the ordered apple juice to, perhaps as one example, turn disordered cytoplasmic ADP into stored, ordered ATP,\cite{note24} using cellular mitochondria as the tool. This is not just a {\it static} chemical reaction, it is {\it not thermostatics}, 
but an {\it ongoing} ordering process driven the disordering arrow in time 
feeding on the ordered food, and the {\it answer} to the Big Question 
raised by Schr\"odinger and Nurse. All in full compliance with the arrow 
in time and the order creation principle. As we said earlier:  
{\it Those naturally cellular tools, empowered by the disordering food in 
the cellular cytoplasm, perform the ongoing ordering tasks necessary 
to continuously sustain the ordered cell, {\it i.e.} perform the 
fundamental task of homeostasis -- it is thermodynamics 2.0.} 

\subsubsection{Exploiting the kinetic order in sunlight} 
Yet there remains a fly in the ointment. Where does the ``ordered food, 
the apple juice'' that is absorbed into the cellular cytoplasm, come from in the first place? It appears that only order that life has to feed upon is other forms of ordered life! Life feeds on life.

Surely you see the design flaw in that plan. Life cannot exist by feeding entirely upon itself, thereby creating the mythical ouroboros. To do so there would continue to be a net worldwide disordering, and life, the entirety of all life everywhere, would perish for, as Schr\"odinger said in his version of ``What is 
life?'': ``Thus, a living organism [and the entirety of life] continually 
increases its entropy [disorder] ... and thus tends to approach the 
dangerous state of maximum entropy, which is [universal] death.'' 
There must, therefore, be some source of order outside the body of all 
life on Earth if life on Earth is to survive! 

Of course, we know what that outside source is; it is our magnanimous sun, Sol, the greatest of all (semi-local) macroscopic systems with a ready access to macrostates of greater disorder; it does so by shedding ordered sunlight the live-long day. 

But this form of order, this kinetic order in the sunlight reaching Earth, is order of a different kind. It is not the spatial streaming order in sunlight or atmospheric winds in which all the particles are moving in the same direction. But it is clearly ordered since there is a more disordered state available, to wit:

Those high energy solar photons have a frequency and a density that is 
perfect for photosynthesizes in plants. That photon stream has cooled 
in the journey from the sun to the Earth, not by lowering the photon 
frequency/energy, but by lowering the photon density in the $1/r^2$ dispersion, $r$ being 
the distance between the Sun and the Earth. Photosynthesis requires 
photons with sufficiently high energy to strip tightly-bound electrons 
from the chlorophyll molecule (or silicon in solar cells) that reside 
in the various shades of green plants, but, critically, not fry the 
plant due to their overall intensity, 
{\it i.e.} heat. Nor is absorbing a succession of low energy photons an 
option, as Einstein pointed out in that year {\it mirabilis} 1905. 
There is a quantum of binding energy in the chlorophyl’s bound electron, 
and it is not to be stripped away by anything less than a photon of 
sufficient energy. {\it And that is provided by the kinetic order in 
sunlight,\cite{note25} a sufficient spectrum of high energy photons in 
the cool environment of the fragile plants on Earth.} The is kinetic 
order in sunlight reaching the Earth.

\subsubsection{An even briefer history of time}
Some may find majesty in the breathtaking immensity of the cosmos, or the prospect of individual particles traveling through wormholes in time, but for me it is the irony of this grand spectacle of life in which it is first disordered (cellular entropy/disorder increases) as we breathe, think and play, in short anabolize, and then reordered (the cellular entropy/disorder decreases) as we eat, drink and, in short, metabolize food. Anabolism (the natural dissipative functions) followed by catabolism (the natural restorative functions,) {\it both 
disordering and reordering, each directionally empowered by the autonomous, 
disordering arrow in time!} That’s the stunning, dumbfounding, 
inspirational story of life! Vive Life!

\section{Summary}
In the end, however, life is just a manifestation of the curious way in which the universe has followed the universal arrow in time after the Big Bang projected the universe toward ever increasing entropy/disorder. The astonishing origin of life follows our order creation principle. We might speculate:  
{\it Once upon a time there was this two component system: one component the ordered sunlight, and, as a second component, a chemical composition of oxygen, hydrogen, nitrogen, carbon, calcium, and phosphorus atoms. The two working together to provide the foundation for life. The sunlight gave up a fraction of it kinetic order to the lesser positional ordering of the chemical composition, to create the first self-replicating compound, and off it went along the path of natural selection in species creation.} \\
I don't think it's possible to come up with a briefer history of 
time -- or one more lacking in critical detail.

In the end, however, there will be a {\it net} disordering of the whole. 
This brief interlude of ordered life in the universe will turn out to be 
just one grand tease! If there are to be multiple advanced civilizations, 
it would be interesting to compare them, if only to see how their 
gods differed. 	



\begin{thebibliography}{27}
\bibitem{note1} But, as a measure of that difference in emphasis between these 
two books on that “big question,” we note that Schr\"odinger’s book 
employed the word “entropy” fifty times (counted from the Kindle version) 
but it appears not at all 
in Nurse's book, though Nurse does speak of ``disorder'' twice in the sense in 
which Schr\"odinger speaks of “entropy.”
\bibitem{note3}
  As we have said, it's not a consequence of the second law, a static law, but the dynamic arrow in time that, statistically, propels macroscopic matter to ever greater entropy/disorder.
\bibitem{note5}
  It is possible to derive a corollary to the second law that says: when a constraint is relaxed on a macroscopic system at equilibrium, the entropy of the any future equilibrium state after must be greater than or equal to that of the original equilibrium state, but says nothing about the process. 
Proof: simple. First, the entropy/disorder/probability of the original equilibrium state is as large or larger than that of any other macrostate compatible with the existing macrostate; that’s the law, the second law of thermodynamics. Now, if we relax a constraint on that original system and allow the system to come to new equilibrium state, the new entropy/disorder must again be a maximum with respect to any accessible state. But, since it was only the relaxing of a constraint, the system continues to have access to the original state, and so the new entropy must be as great or greater than the original entropy -- QED.
\bibitem{note8}
  As a trivial example, take a box containing two pennies. Initially they are both heads. Shake the box. More often than not the result will be one head and one tail, for there are two ways to get one head and one tail, but only one way to get either two heads or two tails. If the number of pennies were on the order of Avogadro's number, $6.02\times 10^{23}$, a typical macroscopic system, the variance from the more probable state of half heads and half tails would be, though the wonders of crowd behavior and the law of large numbers, on the order of $10^{-12}$.
\bibitem{note10}  A snowball in the arctic is without the access of the iconic snowball in hell.
\bibitem{note11}
Nobody really understands energy anyway. Here is Richard Feynman on the subject:
``It is important to realize that in physics today, we have no knowledge of 
what energy is. We do not have a picture that energy comes in little blobs 
of a definite amount. It is not that way. However, there are formulas for 
calculating some numerical quantity, and when we add it all together it gives 
28 [for example] -- always the same number.''
\bibitem{note12}
 If $S$ is a maximum at equilibrium, then it is a straightforward matter to 
show that $dS=dQ/T$. That equality provides a bridge between the thermostatics of Clausius, in which $dS=dQ/T$ is a definition, and the Boltzmann/Gibbs thermostatics in which that equation may be deduced from the deeper principle that entropy is a measure of the probability, and is therefore a maximum at equilibrium.
\bibitem{note14}
  It may seem that there are as many of one as the other, as many disordering  trajectories as ordering, since for every trajectory evolving  one way in time, there is another, with velocities reversed, evolving the opposite way in time. To understand the fallacy in the this argument, see Hurley(2019).\cite{note7}
\bibitem{note16}
In the earlier example of heat exchange between two bodies, one hot and one cold, that continuous trajectory toward a common equilibrium temperature is one example, though we cannot follow the details of the disordering.
\bibitem{note22}
  Water at elevation is more ordered than the same water, at the same temperature, at ground level. Just pull the plug to witness the autonomous flow of
elevated water to the ground.
\bibitem{note23}
  To identify precisely how these tools produce the desired ordering contribution to the cell is a prodigious task for the chemist; witness the Krebs cycle in mitochondria. The Krebs cycle represents the traditional continuous disordering of the arrow time, and the resulting equilibrium product the traditional stability of the eventual equilibrium state.
\bibitem{note24}
  ATP is often -- generally -- referred as the energy powerhouse of the cell. It is nothing of the sort! The value of ATP is its chemical affinity for change, its ready access, for turning into ADP and, in the process, performing some useful function within the cell. 
Why not use food to directly perform the task, you ask? Why create ATP as the intermediary? Food may be willing and able, but ATP is willlinger still! Ask any qualified chemist.
\bibitem{note25}
  Beside the kinetic order in sunlight (high energy photons in the cool environment of the Earth) there is also the directional order in sunlight that we spoke of earlier in the generation of ordered winds in the atmosphere.
\end{thebibliography}
\end{document}